\documentclass[12pt]{article}%
\usepackage{amsmath,bm}
\usepackage{amsfonts}
\usepackage{amssymb}
\usepackage{graphicx}
\usepackage{rotating}
\usepackage[breaklinks=true,bookmarksopen=true,colorlinks=true,citecolor=blue]%
{hyperref}
\usepackage[authoryear]{natbib}
\usepackage{color}
\usepackage{colortbl}
\usepackage{paralist}
\usepackage{amsmath}%
\usepackage[utf8]{inputenc}
\usepackage{amsmath}
\usepackage{tabularx}
\usepackage{makecell}
\usepackage{natbib} 
\usepackage[a4paper, margin=1in]{geometry}
\usepackage{setspace}
\usepackage{caption} 
\usepackage{graphicx}

\begin{document}

\title{Panel Estimation of Taxable Income Elasticities with Heterogeneity and Endogenous Budget Sets\thanks{soren.blomquist@nek.uu.se, anil-kumar@uiowa.edu, wnewey@mit.edu

This research was supported by NSF Grant 224247. Helpful comments were provided by P. Kline, R. Matzkin, and participants at the 2022 Berkeley Conference in Honor of James L. Powell.}}
\author{Soren Blomquist\\\textit{Uppsala University}
\and Anil Kumar\\\textit{University of Iowa}
\and Whitney K. Newey\\\textit{MIT and NBER}}
\date{December 2024}
\maketitle

\begin{abstract}

This paper introduces an estimator for the average of heterogeneous elasticities of taxable income (ETI), addressing key econometric challenges posed by nonlinear budget sets. Building on an isoelastic utility framework, we derive a linear-in-logs taxable income specification that incorporates the entire budget set while allowing for individual-specific ETI and productivity growth. To account for endogenous budget sets, we employ panel data and estimate individual-specific ridge regressions, constructing a debiased average of ridge coefficients to obtain the average ETI. Using data from the Panel Study of Income Dynamics (PSID) spanning 1977–1997, we estimate an average ETI of 0.605.

\end{abstract}

\noindent \textbf{Keywords:} Panel data, Endogenous nonlinear budget sets, Taxable income, Heterogeneous preferences, Heterogenous productivity growth, 
Debiased average ridge estimator
\\
\noindent \textbf{JEL Classification:} C14, C23, H24, H31, J22

\newpage
\doublespacing
\section{Introduction}
\label{sec:intro}
Behavioral responses to tax changes are of significant policy interest, and understanding these responses is crucial for designing effective tax policies. These responses are often summarized by two key measures: the taxable income elasticity, which captures how taxable income reacts to changes in the slope of the budget constraint, and the income effect, which reflects how taxable income responds to a vertical shift in the budget constraint. These summary measures offer valuable insights into the shape of an optimal income tax schedule and the potential impact of tax reforms on taxable income. However, given that tax schedules are typically non-linear and often involve piece-wise linear structures, with reforms frequently altering kink points, virtual incomes, and segment slopes, a more comprehensive approach is needed. 
Specifically, a taxable income function that considers the entire budget constraint may be more effective in predicting the impact of such reforms. Allowing for individual heterogeneity in preferences and productivity is also potentially important. 
Evidence of heterogeneous taxable income elasticities was provided by Kumar and Liang (2020). 
It is not clear how to interpret estimates that ignore such heterogeneity. 
Also, budget sets may be endogenous. 
In the U.S., the deductability of mortgage interest and property taxes can lead to endogeneity. 

In this paper, we estimate a taxable income function with taxable labor income as the dependent variable, where the regressors depend on the entire budget set.
Building on an isoelastic utility framework, we derive a linear-in-logs taxable income specification that incorporates the entire budget set.
We allow for individual heterogeneity while also controlling for budget set endogeneity by using panel data. 
We assume preferences and productivity are individual specific while being drawn from the same distribution in each time period, conditional on the budget sets for all time periods.
This assumption allows individual preferences to be correlated with budget sets with stationarity over time identifying the individual specific parameters. 
In this setting, individual specific regressions are used to estimate preference parameters for each individual.
These estimates are then averaged across individuals to estimate average preferences.  

To allow for nonidentification of parameters for some individuals we use ridge regression for each individual and then debias the average of ridge estimates, as in Chernozhukov et al. (2024). 
The debiased average ridge estimator has an empirical Bayes interpretation, is unbiased if slopes do not vary with individuals, and approaches fixed effects as the ridge regularization grows. 
Standard errors are based on simple formulae.
In addition we describe how to empirically check the extent of identification by comparing coefficients of interest with ridge regularized versions. 

We apply our method to an unbalanced panel dataset from the Panel Study of Income Dynamics (PSID), covering the years 1977 to 1997. Our results yield an average ETI estimate of 0.605. We also find that only a few individuals have parameters that are not well identified.

In a prior study Blomquist et al. (2024), building on Blomquist and Newey (2002), we developed a nonparametric approach to estimate a taxable income function capable of predicting the effects of tax reforms. 
That method, which treats the entire budget constraint as regressors, also allows for multidimensional preference heterogeneity and heterogeneous ETI, while accounting for exogenous productivity growth. 
However, that study required the budget sets to be statistically independent of preferences and that exogenous productivity growth was uniform across years and individuals.
The results we give here address these issues by using panel data to allow for endogenous budget sets and heterogeneous productivity growth.

The advantages of our approach come with certain trade-offs. 
Here we adopt a particular functional form for the utility function that yields a taxable income function similar to those used by Gruber and Saez (2002) and Blomquist and Selin (2010). 
As such, our method is semi-parametric.
We innovate in specifying a model that allows for individual specific taxable income elasticities and productivity growth rates. 

The study of behavioral responses to tax changes has long been a central area of economic research. Historically, much of the focus was on labor supply, with the primary question being how labor supply responds to tax reforms. In a series of influential papers, Feldstein (1995, 1999) argued that individuals respond to tax changes on multiple margins beyond hours worked. These include exerting more effort in current employment, switching to higher-paying jobs that require more effort, or relocating geographically to better-paying positions. Other margins include choosing different compensation mixes (e.g., cash vs. fringe benefits) and engaging in tax avoidance or evasion. By estimating how taxable income responds to changes in the marginal net-of-tax rate, one can capture a broader set of these relevant margins. Feldstein's empirical work found a taxable income elasticity of 3. Following Feldstein's work, a large body of literature emerged, producing a wide range of elasticity estimates, from -1.3 (Goolsbee, 1999) to 3 (Feldstein, 1995). More recent studies, including Saez (2003), Gruber and Saez (2002), Kopczuk (2005), and Giertz (2007), produced estimates closer to 0.5, with Saez, Slemrod, and Giertz (2012) providing a comprehensive review of the literature up to 2012. Subsequent studies by Weber (2014), Burns and Ziliak (2017), and Kumar and Liang (2020) have reported higher elasticity estimates ranging from 0.6 to 1.4. While these conventional elasticity estimates are useful for understanding how taxable income responds to marginal changes in a linear budget constraint, they are less effective in predicting the impact of tax reforms on taxable income. In reality, tax systems are non-linear, and reforms often result in changes to both kink points and marginal tax rates across different income brackets.

\section{Theoretical Framework}

With the exception of Blomquist et al. (2024), all previous studies of the taxable income elasticity have used a parametric taxable income function. Often, the estimated taxable income function has had the form \(\ln Y = \theta \ln \rho + \ln \eta\), where \(Y\) is the taxable income, \(\rho\) the slope of a linear budget constraint, \(\theta\) the taxable income elasticity, and \(\eta\) represents unobserved preference heterogeneity, sometimes interpreted as innate ability. This equation results from maximizing the isoelastic utility function subject to a linear budget constraint. The isoelastic utility function is:
\[
U(C, Y, \theta, \eta) = C - \frac{\eta}{1 + \frac{1}{\theta}} \left( \frac{Y}{\eta} \right)^{1 + \frac{1}{\theta}} \tag{1};\eta>0,\theta>0.
\]
where \(C\) is consumption. 
Note that this utility function is linear in \(C\), implying there are no effects of non-labor income on taxable income.
Maximizing this utility function subject to a linear budget set with slope $\rho$ gives $ln(Y)=\theta*ln(\rho) + \eta$.

We adopt this utility function and assume that  \(\theta\) is an individual specific parameter and integrate over \(\eta\) to obtain a budget set regression. 
For each individual \(i\) the parameter \(\theta_i\) is individual specific as is the distribution of \(\eta_{it}\). 
An identifying assumption is that for each individual the distribution of \(\eta_{it}\) does not vary with $t$.
Consequently, \(\theta_{i}\) may be identified from variation over time in budget sets.
This kind of specification is common in nonlinear panel models such as discrete choice; see Chamberlain (1984).
We innovate in specifying a budget set regression, rather than a discrete choice model, and in allowing a slope parameter \(\theta_{i}\) to vary by individual.

To describe the budget set regression let \(\eta_{it}\) have a probability density function \(g_i(\eta)\). 
Assume that the budget constraint is piecewise linear and convex consisting of \(J\) segments and \(J-1\) kink points. 
Denote the slope and right kink of segment \(j\) as \(\rho_j\) and \(K_j\) respectively and let \(\tilde{Y} = \ln Y\), \(\tilde{\rho_j} = \ln \rho_j\), and \(\tilde{l}_j = \ln K_j\).
Then similarly to Blomquist and Newey (2002) we can integrate over $\eta_{it}$ to get
\[
E(\tilde{Y}_{it} | \text{budget set}_i) = a_i + \theta_i \tilde{\rho}_{Jit} + \sum_{j=1}^{J_{it}-1} \left[ \xi_i(\tilde{l}_{jit} - \theta_i \tilde{\rho}_{jit}) - \xi_i(\tilde{l}_{jit} - \theta_i \tilde{\rho}_{j+1,it}) \right] \tag{3}
\]
for an individual specific intercept $a_i$ and a function $\xi_i(v)$. 
In this paper, we assume \(\xi_i\) is linear: \(\xi_i(v) = \bar{b}_i + \bar{c}_i v\). Then, equation (3) simplifies to:
\[
E(\tilde{Y}_{it} | \text{budget set}_i) = a_i + \theta_i \tilde{\rho}_{Jit} + c_i (\tilde{\rho}_{Jit} - \tilde{\rho}_{1it}) \tag{4}
\]
where \(c_i = \theta_i \bar{c}_i\). The combination of a linear in logs taxable income function for a linear budget constraint and a linear form for the \(\xi\) function means that many terms cancel, so that the term correcting for the nonlinearity of the budget constraint is simpler in form as compared to the term in Blomquist and Newey (2002).

It is important to allow for productivity growth that has two effects. First, we must distinguish between taxable income \(Y\), which is the income the government taxes, and the variable in the utility function, which is work effort broadly defined, denoted as \(\breve{Y}\). The two variables are related by \(Y = \varphi(t) \breve{Y}\), where \(\varphi(t)\) is exogenous productivity growth. We normalize \(\varphi(0) = 1\) for some period \(t = 0\). Second, the slope \(\rho\) indicating how much more consumption the individual receives for a unit increase in \(\breve{Y}\) should be multiplied by \(\varphi(t)\). Hence, we have:
\[
\breve{Y}_t = (\varphi(t) \rho)^\theta \eta \tag{5}
\]
Multiplying \(\breve{Y}\) by \(\varphi(t)\), we get:
\[
Y = \varphi(t) \breve{Y} = \varphi(t) (\varphi(t) \rho)^\theta \eta \tag{6}
\]
Taking logs and using the previous notation, we obtain:
\[
\tilde{Y} = (1 + \theta) \ln \varphi(t) + \theta \tilde{\rho} + \tilde{\eta} \tag{7}
\]
This is a form that is commonly estimated. One does not try to identify \(\theta\) from the term \((1 + \theta) \ln \varphi(t)\), but from \(\theta \tilde{\rho}\). 

The productivity term is often generalized to differ between individuals:
\[
\tilde{Y}_{it} = (1 + \theta_i) \ln \varphi(i,t) + \theta_i \tilde{\rho}_{it} + \tilde{\eta}_{it} \tag{8}
\]
Often, \(\varphi(i,t)\) is modeled as a function of characteristics like birth cohort, state, or region (see, for example, Burns and Ziliak, 2017). A convenient feature of equation (7) is that the productivity term enters additively. Panel data is commonly used to estimate this function. In this paper, we assume:
\[
\varphi_i(t) = e^{\alpha_i(t - 1)} \tag{9}
\]
The fixed effects regression for a convex, piecewise linear budget constraint then becomes:
\[
E(\tilde{Y}_{it} | \text{budget set}) = \hat{a}_i + \theta_i \tilde{\rho}_{Jit} + c_i (\tilde{\rho}_{Jit} - \tilde{\rho}_{1it}) + d_i t, \quad d_i = \alpha_i(\theta_i + 1) \tag{10}
\]
Here unobserved heterogeneity in productivity is allowed for and not just observed heterogeneity. 

We also estimate a function that includes an income effect. For a linear budget constraint with slope \(\rho\), lump-sum income \(R\), and ignoring productivity growth, the function we estimate is:

\[
Y = \kappa \rho^\theta R^\gamma
\]
or in logarithmic form:
\[
\ln Y = \ln \kappa + \theta \ln \rho + \gamma \ln R \tag{11}
\]
Burtless and Hausman (1978) used this functional form to estimate a labor supply function.  They also derived the corresponding indirect utility function and gave necessary and sufficient conditions for this function to be consistent with utility maximization. Sufficient conditions for the Slutsky condition to be satisfied are \( \theta \geq 0\) and \(\gamma \leq 0\). In a footnote they mention that the direct utility function can be derived from the indirect utility function, but that a closed form solution does not exist.  Stern (1986) provides an expression for the direct utility function (it is a non-closed form function). Later the functional form (11) was used by Gruber and Saez (2002) and Blomquist and Selin (2010) to estimate taxable income functions.

When the taxable income function for a linear budget constraint takes the form (11), expressions (3) and (10) must be modified. Denoting the virtual income for segment \(j\) as \(R_j\) and \(\tilde{R} = \ln R\), and ignoring productivity growth, the expected taxable income for a convex, piecewise linear budget constraint becomes:

\begin{align}
E(\tilde{Y}_t | \text{budget set}) = & \; a_i + \theta_i \tilde{\rho}_{Jit} + \gamma_i \tilde{R}_{Jit} + \sum_{j=1}^{J-1} \big[\xi_i(\tilde{l}_{jit} - \theta_i \tilde{\rho}_{jit} - \gamma_i \tilde{R}_{jit}) \nonumber \\
& - \xi_i(\tilde{l}_{jit} - \theta_i \tilde{\rho}_{j+1,it} - \gamma_i \tilde{R}_{j+1,it})\big] \tag{12}
\end{align}

Accounting for productivity growth and assuming \(\xi_i\) is linear, \(\xi_i(v) = \bar{b}_i + \bar{c}_i v\), this simplifies to:
\[
E(\tilde{Y}_{it} | \text{budget set}) = \tilde{a}_i + \theta_i \tilde{\rho}_{Jit} + \gamma_i \tilde{R}_{Jit} + c_i (\tilde{\rho}_{Jit} - \tilde{\rho}_{1it}) + \hat{c}_i (\tilde{R}_{Jit} - \tilde{R}_{1it}) + d_i t \tag{13}
\]
where \(c_i = \theta_i \bar{c}_i\), \(\hat{c}_i = \gamma_i\bar{c}_i\), and \(d_i = \alpha_i (\theta_i + 1)\).

\section{Estimation Framework}

Our estimation framework is based on the debiased average ridge estimator proposed by Chernozhukov et al. (2024), which allows for heterogeneity across individuals in economic models. 
This approach is particularly useful in settings where endogeneity poses a challenge, as it enables the estimation of counterfactual effects when regressors are correlated with unobservables, as is the case with the budget set in a taxable income regression. 
The framework is grounded in a time stationarity assumption for panel data, which states that the conditional distribution of unobserved heterogeneity in any time period given regressors in all periods does not vary with time. 
This assumption enables identification of averages of individual-specific parameters and counterfactual effects.

The primary advantage of this methodology is its ability to handle general multidimensional heterogeneity without assuming additively separable models. 
This is particularly important in economic applications where price effects, income effects, and other factors vary significantly across individuals. 
The framework allows us to estimate the effects of interest under time stationarity of preferences or technology, which, in turn, helps isolate counterfactual effects based on observable variation.

\subsection{Ridge Regression and Debiasing}

To regularize the estimation process and address potential issues of singularity in the second-moment matrix of individual-specific regressors, we apply individual-specific ridge regression. This helps mitigate problems of nonidentification or near singularity in the data. For each individual, we perform ridge regression on the relevant regressors and then debias the resulting coefficients to obtain estimates of average effects.

Let \( \beta_i \) represent the regression coefficients for individual \( i \), which are estimated via ridge regression. The ridge estimator is given by:
\[
\hat{\beta_i} = \left( Q_i + \lambda S_i \right)^{-1} \frac{1}{T} \sum_{t=1}^{T}b_{it}\tilde{Y}_{it}
\]
where $b_{it}$ is the vector of individual specific regressors in time $t$, having $1$ as its first component, $Q_i=\frac{1}{T} \sum_{t=1}^{T}b_{it}b_{it}'$, \( S_i \) is a diagonal matrix that serves as a ridge penalty and has $0$ as the first diagonal element, and \( \lambda \) is a scalar regularization parameter.
The regularization matrix $S_i$ could have $1$ as all of its nonzero diagonal elements. 
Another choice for \( S_i \) is \( S_i = (0, Q_{i22}, \dots, Q_{iJJ}) \), which adjusts the ridge penalization to ensure consistent scaling for each regressor.

\subsection{Debiased Average Ridge Estimator}
We debias the average of these ridge estimators using a matrix adjustment:
\[
W_i = \left( Q_i + \lambda S_i \right)^{-1} Q_i
\]
The debiased average ridge estimator is then
\[
\tilde{\beta} = \bar{W}^{-1} \frac{1}{n} \sum_{i=1}^{n} \hat{\beta_i}, \; \bar{W} = \frac{1}{n} \sum_{i=1}^{n} W_i.
\]
This $\tilde{\beta}$ estimates the average of $\beta_i$ over individuals and thus allows for heterogeneous effects. 
In economic models, where coefficients may vary by individual, this estimates the average of the individual coefficients. 
The $\tilde{\beta}$ can be interpreted as an empirical Bayes estimator of a common prior mean and the ridge regularization parameter \( \lambda \) allows the estimator to vary between the linear fixed-effects estimator (as \( \lambda \to \infty \)) and the average of individual-specific least squares estimates (as \( \lambda \to 0 \)); see Chernozhukov et al. (2024) for these properties.

The asymptotic variance of \( \sqrt{n}(\tilde{\beta} - \beta_0) \) is estimated as follows:
\[
\hat{V} = \frac{1}{n} \sum_{i=1}^{n} \hat{\psi}_i \hat{\psi}_i', \quad \hat{\psi}_i = \bar{W}^{-1} (\hat{\beta}_i - W_i \tilde{\beta}).
\]
An alternative formula for \( \hat{V} \) is:
\[
\hat{V} = \bar{W}^{-1} \frac{1}{n} \sum_{i=1}^{n} (\hat{\beta}_i - W_i \tilde{\beta})(\hat{\beta}_i - W_i \tilde{\beta})' \bar{W}^{-1'}.
\]
In the application we compute the average ridge estimator and a standard error for many values of \( \lambda \). 
Doing this allows us to check the sensitivity of estimates to the degree of ridge regularization.

\section{Data}

The data used in this study come from the Panel Study of Income Dynamics (PSID), covering the years between 1977 and 1997. The PSID is a longitudinal survey that began in 1968 and follows a representative sample of U.S. individuals and their households. To ensure consistency across time, we used the version of the PSID data provided by the Cross-National Equivalent File (PSID-CNEF), which harmonizes key variables across survey waves.

We restricted our analysis to households where the household heads are aged 25 to 60, focusing on prime-age individuals who are typically active in the labor market. Households from the Survey of Economic Opportunity subsample were excluded due to their non-random selection. Additional exclusions removed households with multiple heads or partners in a given year, and we retained only those households with at least 15 years of continuous data.

Our sample covers a period of significant tax reforms in the U.S., including the Economic Recovery Tax Act of 1981 (ERTA) and the Tax Reform Act of 1986 (TRA). We stop at 1997, as the PSID switched to biennial data collection afterward, and restricting the sample to pre-1997 ensures annual data continuity. The final sample includes 1,578 individuals observed for up to 21 years, with at least 15 observations per household.

\sloppy
The PSID-CNEF dataset provides variables capturing household pre- and post-government income, labor income, and various forms of transfer income, such as private transfers, public transfers, and social security income. All income variables were adjusted for inflation using the Consumer Price Index for Urban Consumers (CPI-U) to ensure comparability across years.

The dependent variable is the logarithm of taxable income, represented by the household's labor income. We derived a parsimonious specification based on four key variables to capture the budget set on the right-hand side: the last-segment slope and virtual income, both in logarithms, and their differences from the first-segment slope and virtual income. To calculate these variables, we constructed the complete budget set for each household using the NBER TAXSIM calculator (Feenberg and Coutts, 1993). A range of income levels was run through TAXSIM to obtain federal and state marginal tax rates, which were used to construct the slopes and kink points of the household budget sets. These tax data account for federal and state income taxes, as well as payroll taxes.

Non-labor income, also referred to as the virtual income for the first budget segment, was constructed by combining various components of household income. These include changes in the household’s asset position, post-government income, taxes, imputed rental income from owner-occupied housing, and windfall income, while subtracting labor income and certain tax liabilities. Changes in assets were calculated by capitalizing asset income at a fixed rate of return (assumed to be 6 percent) and taking the difference between consecutive years. This approach provides a comprehensive measure of non-labor income available to households, incorporating income from assets, housing, and government transfers.

To compute the virtual income for subsequent budget segments, the virtual income of each segment was derived by adjusting the virtual income of the previous segment using the difference in slopes between consecutive segments and the kink points.

The final dataset is well-suited for examining the effects of tax policy changes over time, offering sufficient variation in income and tax rates, driven by tax reforms and household characteristics. Descriptive statistics for key variables are presented in Appendix Table 1.

\section{Results}

\begin{table}[h!]
\centering
\caption{Slope Elasticity Estimates Without Income Effect}
\vspace{\baselineskip}
\resizebox{\textwidth}{!}{
\begin{tabular}{cccc} 
\hline
\hline
\textbf{Lambda} & \textbf{Non-debiased Slope Elasticity} & \textbf{Debiased Slope Elasticity} & \textbf{Standard Error} \\
\hline
0 & 0.2022963 & 0.2027358 & 0.2446697 \\
1.00e-07 & 0.2032239 & 0.203854 & 0.2407696 \\
1.00e-06 & 0.2075759 & 0.2084525 & 0.2187759 \\
1.00e-05 & 0.161933 & 0.1630704 & 0.1582035 \\
.0001 & 0.1361381 & 0.1367891 & 0.1182584 \\
.001 & 0.2052194 & 0.2541288 & 0.0931669 \\
.01 & 0.1830269 & 0.4315851 & 0.0818147 \\
.1 & 0.1010592 & 0.623476 & 0.0815514 \\
.2 & 0.0752519 & 0.6952185 & 0.0852087 \\
.3 & 0.0607792 & 0.7349355 & 0.088131 \\
.4 & 0.0511856 & 0.760554 & 0.0903812 \\
.5 & 0.0442829 & 0.7785257 & 0.0921443 \\
1 & 0.0265913 & 0.8224351 & 0.0971883 \\
2 & 0.0148325 & 0.8495471 & 0.1009638 \\
3 & 0.0102909 & 0.8594759 & 0.1025208 \\
\hline
\hline
\end{tabular}
}
\label{tab:Table1}
\end{table}

Tables \ref{tab:Table1} and \ref{tab:Table2} present the estimates of the average ETI  across various values of the regularization parameter \(\lambda\). Both non-debiased and debiased estimates are reported, along with their corresponding standard errors. The debiasing procedure corrects for the regularization bias in ridge regression, providing more accurate estimates of the average ETI. The debiased estimate of the average ETI is of primary interest in this analysis.

The results in Table \ref{tab:Table1} show a stark contrast between the non-debiased and debiased average ETI estimates. 
The non-debiased estimates shrink in magnitude as $\lambda$ increases, exhibiting shrinkage bias from ridge regularization. 
The debiased estimates exhibit first shrink and then increase as \(\lambda\) increases. 
This divergence between average ridge and debiased estimates highlights the importance of debiasing. 

Focusing on the debiased estimator we observe that as \(\lambda\) increases from zero the average ETI estimate becomes significant at  \(\lambda = 0.001\), and remains significant as \(\lambda\)  increases further. At \(\lambda = 0.001\) the average ETI estimate reaches a notable value of 0.254 with a standard error of 0.093. This choice of lambda represents a balance between regularization and statistical significance. 
Beyond this point, as \(\lambda\) increases, the debiased average ETI continues to rise although that could result from over-regularization.

\begin{table}[h!]
\centering
\vspace{1em} 
\caption{Results of Non-debiased and Debiased Slope and Income Elasticities with Standard Errors}
\vspace{\baselineskip}
\small  
\resizebox{\textwidth}{!}{
\begin{tabular}{ccccccc} 
\hline
\hline
\textbf{Lambda} & \makecell{\textbf{Non-debiased} \\ \textbf{slope} \\ \textbf{elasticity}} & \makecell{\textbf{Debiased} \\ \textbf{slope} \\ \textbf{elasticity}} & \makecell{\textbf{Standard} \\ \textbf{Error}} & \makecell{\textbf{Non-debiased} \\ \textbf{income} \\ \textbf{elasticity}} & \makecell{\textbf{Debiased} \\ \textbf{income} \\ \textbf{elasticity}} & \makecell{\textbf{Standard} \\ \textbf{Error}} \\
\hline
0          & 0.780617 & 0.787642 & 0.415832 & 0.01345  & 0.007462 & 0.062251 \\
1.00E-07   & 0.677571 & 0.685251 & 0.37199  & 0.001554 & 0.001543 & 0.053651 \\
1.00E-06   & 0.597343 & 0.605638 & 0.288264 & 0.007321 & 0.007399 & 0.035524 \\
1.00E-05   & 0.474162 & 0.486876 & 0.205531 & 0.026225 & 0.027226 & 0.02018  \\
0.0001     & 0.34046  & 0.368166 & 0.148268 & 0.016726 & 0.020582 & 0.014406 \\
0.001      & 0.295437 & 0.3787   & 0.107041 & -0.0011  & 0.001402 & 0.010159 \\
0.01       & 0.220099 & 0.531998 & 0.093561 & -0.00267 & -0.00328 & 0.009117 \\
0.1        & 0.09549  & 0.759263 & 0.091503 & -0.00083 & 0.003373 & 0.009558 \\
0.2        & 0.067323 & 0.847766 & 0.094317 & -0.00053 & 0.006233 & 0.009713 \\
0.3        & 0.053496 & 0.898555 & 0.097036 & -0.00041 & 0.007672 & 0.009782 \\
0.4        & 0.044842 & 0.93253  & 0.099342 & -0.00034 & 0.008516 & 0.009827 \\
0.5        & 0.038785 & 0.957175 & 0.10128  & -0.00029 & 0.009057 & 0.009861 \\
1          & 0.023564 & 1.021592 & 0.107563 & -0.00018 & 0.010123 & 0.009993 \\
2          & 0.013386 & 1.066017 & 0.113199 & -0.0001  & 0.010501 & 0.010166 \\
3          & 0.009376 & 1.083669 & 0.115836 & -7.2E-05 & 0.010553 & 0.010268 \\
\hline
\hline
\end{tabular}
}
\label{tab:Table2}
\end{table}

Turning to Table 2, which incorporates the income effect, similar results are observed. The non-debiased slope average ETI estimate declines monotonically with $\lambda$. 
The debiased estimate initially declines but increase as \(\lambda\) rises, peaking at 1.08 for \(\lambda = 3\). 
The debiased estimate becomes statistically significant at \(\lambda = 1.00E-06\), where it is 0.605 with a standard error of 0.288.

As mentioned in Section 3.2, the debiased average ridge estimator varies between an average of individual slopes and the fixed effect estimates as $\lambda$ varies between zero and one. 
Thus, when there is individual heterogeneity in true slopes we should expect our estimates to vary with $\lambda$ as they do in Table 2.
That variation is consistent with the presence of individual heterogeneity in the ETI.

For \(\lambda = 1.00E-06\),  the estimated nonlabor income elasticity is 0.0074 with a standard error of 0.0355. This implies a 95\% confidence interval of (-0.0622, 0.0770). The estimate is neither significantly different from zero nor from -0.06. Likewise, if we define the nonlabor income elasticity as (dY/dR)(R/Y), and use the fact that  Y/R is approximately 5.96 , the confidence interval for the nonlabor income effect would be (-0.371,0.459). This finding, that the income effect is estimated with a large standard error, aligns with many other results for panel data studies of taxable income.

One reason it is hard to estimate the effect of the nonlabor income with precision is that nonlabor income tends to change little from year to year for a given individual. 
Thus, panel data normally used are not well designed to accurately capture the nonlabor income effect. Since sizeable precise estimates of nonlabor income effects are rare, many studies neglect to account for nonlabor income, arguing that it is known that the nonlabor income effect is small. However, this reasoning is at odds with the findings in Golosov et al. (2024). They purposely design a data set that should be able to detect a nonlabor income effect, if one exists.

Utilizing annuitized lottery data- taking into account that a large lottery winning or inheritance can be spread out over many years by savings or dissavings- they find a substantial effect of exogenous nonlabor income. They report, “On average, an extra dollar of unearned income in a given period reduces household labor earnings by about 50 cents, decreases total labor taxes by 10 cents, and increases consumption expenditure by 60 cents.” Although their data were designed to provide a precise estimate of the income effect, it does not yield a slope elasticity. In future work we aim to construct data that can provide credible estimates of both the slope elasticity and the nonlabor income effect.

 Understanding the nonlabor income effect is just as important as having a reliable estimate of the slope elasticity. First, if we want to predict the effect of tax reforms, say the introduction of a liveable guaranteed income, it would make a large difference whether the nonlabor income effect is zero or say -0.5, which is the estimate obtained in Golosov et al. 
 Second, knowing the nonlabor income effect is important when calculating the compensated taxable income elasticity, which is the relevant elasticity when calculating deadweight losses of taxes.

 Interestingly, although the estimated nonlabor income effect in table two is very imprecise, including nonlabor income has a large impact on the estimated average ETI. 
 In table 1 the average ETI estimate is 0.254 at the smallest $\lambda$ where the estimate is significant and in table 2 it is 0.605.
 Thus the inclusion of nonlabor income leads to increased estimates of the average ETI.

\begin{figure}[htp]
\centering
\includegraphics[scale=0.95]{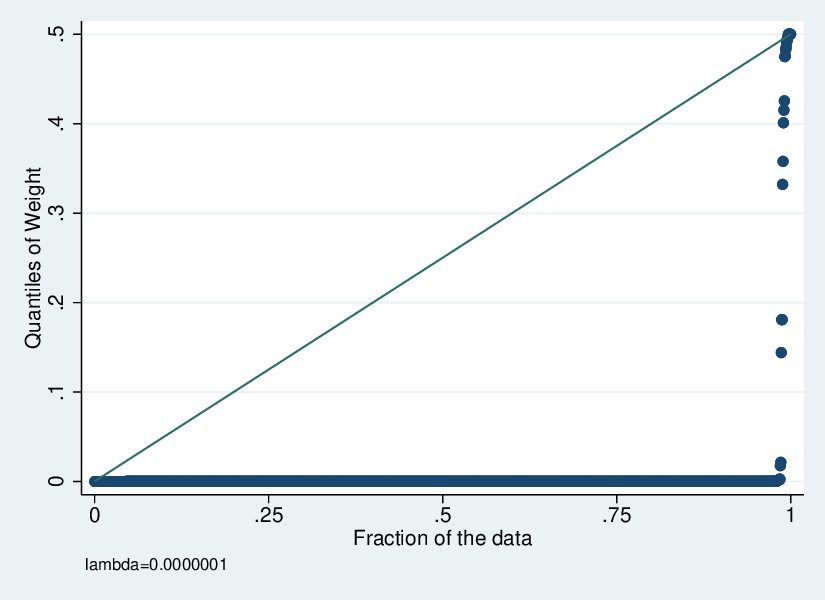}
\caption{Quantile Plot of Regularization Effects}
\end{figure}

\subsection{Identification} 

Identification of the average ETI $\theta_0=E[\theta_i]=E[e_{\theta}'\beta_i]$ depends on how informative each observation $i$ is for $\theta_i$, where $e_{\theta}$ is the unit vector that selects $\theta_i$ from the individual specific coefficient vector $\beta_i$.  
One way to measure this informativeness is to compare $e_{\theta}$ with a regularized version $\hat{e}_{{\theta}i}=e_{\theta}'{\bar{W}}^{-1}W_i$ that determines the contribution of $\beta_i$ to the mean $E[\hat\theta]= \sum_{i=1}^{n} e_{\theta}'{\bar{W}}^{-1}W_i\beta_i/n$ of the debiased ridge estimator. Figure 1 gives a quantile plot of
\[
\zeta_i =\frac{\|\hat{e}_{{\theta}i}-e_{\theta}\|} {\sqrt{2\|\hat{e}_{{\theta}i}\|^2+2}} 
\]
for $\lambda=.0000001$. In this plot, the $\zeta_i$ values quantify the departure of the regularized vector $\hat{e}_{{\theta}i}$ from the unit vector $e_{\theta}$.
Observations with $\zeta_i$ closer to 0 (in the lower quantiles of the distribution) exhibit less regularization and so higher identification strength. Conversely, higher $\zeta_i$ correspond to observations where more shrinkage is applied due to \( Q_i \) being closer to having a null space that includes $e_{\theta}$.
We see that identification is a potential problem for only relatively few observations in this empirical setting.

\section{Conclusion}
\label{sec:conclusion}

This paper introduces a novel approach for estimating the average of heterogeneous elasticities of taxable income in the presence of endogenous, nonlinear budget constraints and individual specific productivity growth using panel data. 
We have used an isoelastic utility function to specify a panel regression that incorporates the entire budget set.
We utilize individual-specific ridge regressions to allow for weak identification and debias the average ridge elasticity estimator. 
In our preferred specification, which includes an income effect, we observe that as the ridge penalty (\(\lambda\)) increases, the average ETI estimate is significant at \(\lambda = 1.00E-06\), estimated at 0.605 with a standard error of 0.288. 
This average elasticity measure, which we identify as the preferred estimate, suggests a robust behavioral response to tax rate changes while accounting for individual heterogeneity and budget-set endogeneity.

These results enhance our understanding of responses to taxation, offering valuable insights. 
By addressing key sources of bias in ETI estimation and incorporating individual heterogeneity, our method significantly advances the empirical toolkit for analyzing the effect of taxes on taxable income.

\bibliographystyle{apalike}
\nocite{*}
\bibliography{references}

\section*{Appendix}
\addcontentsline{toc}{section}{Appendix} 

\begin{table}[htbp]\centering
\def\sym#1{\ifmmode^{#1}\else\(^{#1}\)\fi}
\captionsetup{labelformat=empty} 
\caption{Appendix Table 1: Summary Statistics} 
\begin{tabular}{l*{1}{cccc}}
\hline\hline
                    &        Mean&          SD&         Min&         Max\\
\hline
Taxable Labor Income                   &    59269.53&    50442.83&    .9722068&     1557509\\
First Segment Slope &    1.020727&    .1087607&         .24&         1.5\\
First Segment Virtual Income&    9941.909&    23699.26&    .0005321&     1445785\\
Last Segment Slope  &    .5286873&     .102572&         .25&         .72\\
Last Segment Virtual Income&    32855.45&    29008.94&    .2061627&     1456541\\
Log labor income    &    10.73526&    .8142413&   -.0281867&     14.2586\\
Log last seg. slope &    -.658761&    .2153743&   -1.386294&    -.328504\\
Log last-log first seg. slope&   -.6734313&    .2369308&   -1.425515&    .7137665\\
Log last segment virtual income&    10.08988&    .9518308&   -1.579089&    14.19158\\
Log last-log first segment virtual income&    2.086026&    2.567283&   -8.924937&    17.67892\\
\hline
Observations        &       26107&            &            &            \\
\hline\hline
\multicolumn{5}{l}{\footnotesize Notes: The table presents summary statistics of key variables.}\\
\end{tabular}
\end{table}

\end{document}